# Seismic electric precursors observed prior to the 6.3R EQ of July 1st 2009, Greece and their use in short-term earthquake prediction.

Thanassoulas[1], C., Klentos[2], V., Verveniotis, G.[3], Zymaris, N.[4]

1. Retired from the Institute for Geology and Mineral Exploration (IGME), Geophysical Department, Athens, Greece.
   e-mail: thandin@otenet.gr - URL: www.earthquakeprediction.gr

2. Athens Water Supply & Sewerage Company (EYDAP),
   e-mail: klenvas@mycosmos.gr - URL: www.earthquakeprediction.gr

3. Ass. Director, Physics Teacher at 2nd Senior High School of Pyrgos, Greece.
   e-mail: gver36@otenet.gr - URL: www.earthquakeprediction.gr

4. Retired, Electronic Engineer.

## Abstract.

The normalized raw data of the Earth's electric field, monitored at **PYR, ATH, HIO** monitoring sites in Greece, are studied as far as it concerns the presence of electric seismic precursors. Electric preseismic pulses, plateau-like anomalous electric field and Very Long Period **(VLP)** anomalies were detected prior to the occurrence of the July 1st, 2009 large (Ms = 6.3R) EQ in Greece. Further processing of the same data revealed the true form of the generated preseimic potential thus providing an indirect estimation of the time window of the occurrence of the large EQ. Finally, by correlating, at pairs, the anomalous electric field from all monitoring sites it is possible to calculate the "strange attractor like" seismic electric precursor that provides a time window for the occurrence of the large EQ and utilizes the determination not only of the regional seismogenic area that has been activated but also a more precise epicenter too.

## 1. Introduction.

The concept of the seismic electric precursors, that is the generation of electric signals from the focal area of a large EQ some time prior to its occurrence and their detection at long distances from it, has been extensively speculated in the seismological literature. Papers that range from absolute negation to total acceptance of the concept can be found at a large number for each case respectively (Thanassoulas, 2007). Each presented case is more or less well documented and increases the confusion about the possibility of solving the problem of the EQ prediction in a short-term mode.

In this paper will be presented the seismic electric precursors registered by three monitoring sites **(PYR, ATH, HIO)** located in Greece **(see fig. 1)**, prior to the earthquake of 1st July, 2009 **(Ms = 6.3R).** Moreover, these seismic precursory electric fields will be used for the determination of its epicenter location while time windows of its occurrence will be estimated according to each methodology used.

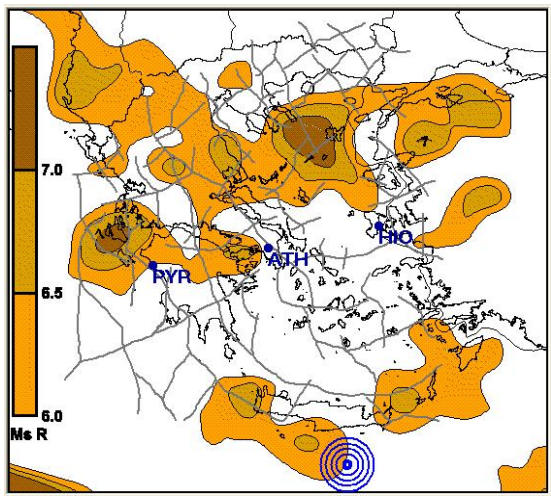

Fig. 1. Location of **PYR, ATH, HIO** monitoring sites in Greece. The blue concentric circles denote the location of the 1st July, 2009, **Ms = 6.3R** EQ. The colored map shows the spatial distribution of the seismic potential (maximum expected EQ magnitude) allover Greece while the gray lines indicate the deep lithospheric fracture zones - faults (Thanassoulas 1998, 2007).

Some technical details of the used array of the monitoring sites are given bellow:
From figure **(1)** the distances between all pairs of monitoring sites and the corresponding EQ have been measured so that one may have an idea about the extend of the used array.
The distances between each monitoring site and the studied EQ are as follows:

   a. **ATH - EQ** = 429.4 Km, b. **PYR - EQ** = 517.6 Km, c. **HIO - EQ** = 448.2 Km.

While the distance between the monitoring sites are as follows:

   b. **ATH - PYR** = 216.6 Km, **ATH - HIO** = 209.5 Km,  **PYR -HIO** = 425.6 Km

The length of the receiving dipoles **(N-S, E-W)** for **PYR** monitoring site equals to **160 m**, for **ATH** monitoring site equals to **21 m** and for **HIO** monitoring site equals to **200 m**.
The recorded raw data have been normalized to equal length of dipoles (at each monitoring site) and to the **N-S, E-W** direction too (Thanassoulas, 2007). All the graphs to follow refer to normalized values of the recorded Earth's electric field.

## 2. Data presentation and analysis.

### 2.1 Normalized raw data.

The data which will be used span from April 10$^{th}$, 2009 to July 15$^{th}$, 2009. It is a total of **86** days. During this period of time only one large EQ took place, the one of July 1$^{st}$, 2009 with a magnitude of **6.3R** in the Richter scale. At first, the normalized data obtained from each monitoring site are presented in the following figures **(2, 3, 4).**

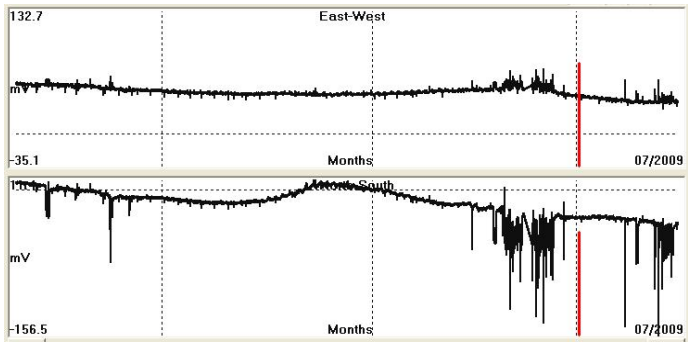

Fig. 2. Normalized raw data observed at **PYR** monitoring site, corresponding to the period 090410 - 090715 (yymmdd format). The red bar indicates the day of occurrence of the large EQ.

The characteristic feature of this recording is the presence of a group of electric spikes, by the end of June, a few days before the large EQ occurrence, in contrast to the previous period of time when no significant electric activity was observed. By comparing the amplitudes of the spikes in both receiving dipoles it is made clear that the azimuthal direction of the intensity vector of the spikes, in general, points towards the forth trigonometric quadrant **(EW=positive, NS=negative)** that is the quadrant, in relation to the **PYR** monitoring site location, in which the large EQ took place. Similar spikes have been reported in the past by Morgounov (2001).
Next, the normalized raw data recorded by the **ATH** are presented in figure **(3).**

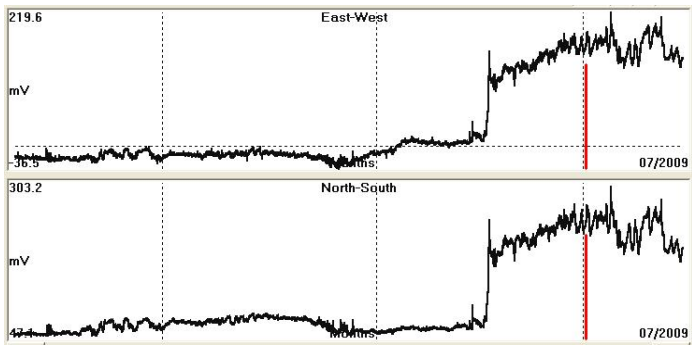

Fig. 3. Normalized raw data observed at **ATH** monitoring site, corresponding to the period 090410 - 090715 (yymmdd format). The red bar indicates the day of occurrence of the large EQ.

The **ATH** monitoring site behaves in an entirely different way. In this case both dipoles register an abrupt positive increase of the registered Earth's electric field. What is interesting in this case is the coincidence of the step like increase time of the electric field to the time of the initiation of the electric spikes registered at **PYR** monitoring site. The electric field increases towards the time of occurrence of the large EQ and afterwards follows a rather irregular form. It represents the typical plateau-like electric anomaly generated by the calculation of the first derivative of an electric field that has been generated by a typical piezoelectric activated mechanism (Thanassoulas, 2008). Considering the calculated azimuthal direction of the Earth's electric field from both **EW** and **NS** components it points towards the first trigonometric quadrant in relation to the location of the **ATH** monitoring site which is irrelevant to the location of the large EQ. A possible explanation of this discrepancy is that the **ATH** monitoring site detects the anomalous electric field but due to the fact that the recording dipoles are very short **(21m)** it is impossible to resolve long wavelengths in a satisfactory way. It rather detects the regional Earth's electric field change, generated by the change of the stress field all over Greece. Indeed, the latter shows this direction **(NE – SW).**
Completing the presentation of the normalized raw data following are the data recorded by the **HIO** monitoring site **(fig. 4).**

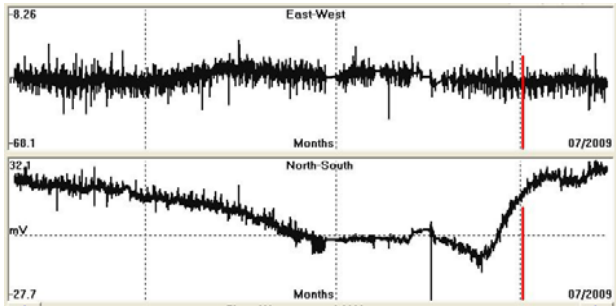

Fig. 4. Normalized raw data observed at **HIO** monitoring site, corresponding to the period 090410 - 090715 (yymmdd format). The red bar indicates the day of occurrence of the large EQ.

The recorded data from **HIO** monitoring site are quite different in form from the **ATH** as well as from **PYR** monitoring sites. The **EW** dipole shows no significantly anomalous electric activity while the **NS** dipole presents the case of a Very Long Period **(VLP)** anomaly. This type of preseismic electric anomaly is very similar in form to the one presented by Sobolev (1975) while the EQ took place at the up-rising branch of it too.

It must be pointed out too that the amplitude of the anomalous electric field, observed at the monitoring sites, varies according to the different type of it. So, the observed electric spikes at **PYR** range from **40mV** to **100 mV** for a dipole length of **160m**. The observed plateau type electric field at **ATH** presents a rising step of **100 – 130mV** for a dipole length of **21m**. The observed **VLP** electric field anomaly at **HIO** ranges from **40mV** to **45 mV** for a dipole of **200m**. The observed amplitude electric field values are related not only to the specific type of recorded anomalous electric field but to the local geological tectonic and stratigraphy conditions too.

## 2.2. T=24h band-pass filtered data.

The already presented data in figures **(2, 3, 4)** are typical cases of time series data which contain a large number of harmonic components. Therefore, the generating (any) mechanism of the recorded data affects, more or less, each harmonic component to a certain extent. Consequently, it is possible to study the evolvement in time of the EQ generating mechanism in a much simpler way if we observe the change of magnitude (and the phase in a lesser extent) in time of a single monochromatic component of the corresponding raw data frequency spectrum. In this case we select and analyze two different harmonic components. The first one corresponds to a harmonic component with **T=24h** (Thanassoulas et al. 1993) while the second corresponds to one of **T=14days**. The choice of these specific components is dictated by the tidal waves deformation effect on the lithosphere caused by the **K1, P1,** for the **T=24h** and **M1** for the **T=14days** (Thanassoulas, 2007).

Both harmonic components **(T=24h, T=14days)** were extracted from the normalized raw data by the use of an appropriate band-pass **FFT** procedure. The obtained results are presented in the following figures **(5, 6, 7)** for **T=24h** and **(8, 9, 10)** for **T=14days**.

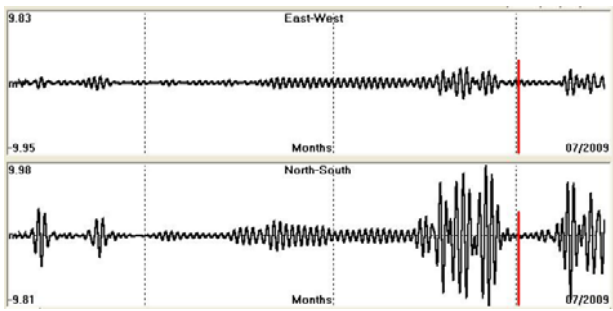

Fig.5. Band-pass filtered **(T=24h)** normalized data from **PYR** monitoring site corresponding to the period 090410 - 090715 (yymmdd format). The red bar indicates the day of occurrence of the large EQ.

The band-pass filtered recorded data at **PYR** monitoring site **(fig. 5)** show an increased in amplitude wavelet of **T=1day** that starts almost fifteen **(15)** days prior to the large EQ occurrence time. Its amplitude drastically decreases a couple of days before the seismic event. The latter complies with the large scale piezoelectric model in conjunction to the lithospheric oscillation triggered by the **K1** and **P1** components of the tidal wave (Thanassoulas et al 1993, Thanassoulas 2007).

Next, the band-pass filtered recorded data at **ATH** monitoring site are presented **(fig. 6).** The initiation of the amplitude increase of the bad-pass filtered data coincides to the one of **PYR** monitoring site shown in figure **(5).** A continuous amplitude increase is observed towards the EQ occurrence time while an amplitude peak is observed concurrently to the seismic event. After the seismic event takes place the amplitude of the oscillation exhibits a random behavior.

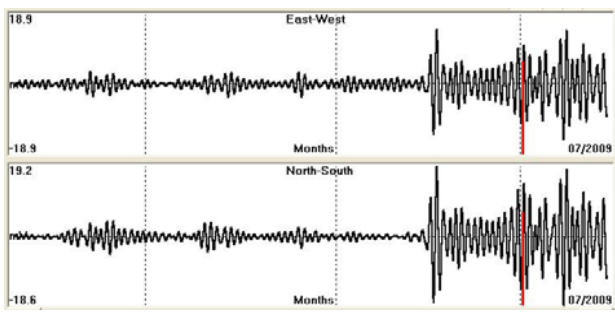

Fig.6. Band-pass filtered **(T=24h)** normalized data from **ATH** monitoring site corresponding to the period 090410 - 090715 (yymmdd format). The red bar indicates the day of occurrence of the large EQ.

In the next figure **(7)** the band-pass filtered recorded data at **HIO** monitoring site are presented. In this case no significant anomalous oscillation amplitude increase was observed.

At all these three monitoring sites **(PYR, ATH, HIO)** and for **T=1day** a quite different behavior was observed. In terms of frequency spectrum content of the recorded electric field, the component of **T=24h** is clearly evident at **PYR** and **ATH** monitoring sites while in the case of the **HIO** monitoring site it coincides more or less to oscillating noise level.

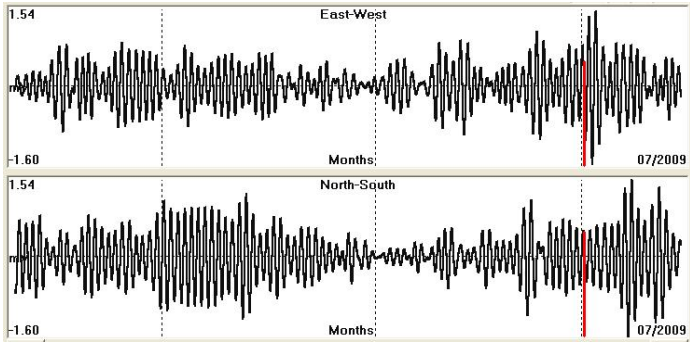

Fig.7. Band-pass filtered **(T=24h)** normalized data **from HIO** monitoring site corresponding to the period 090410 - 090715 (yymmdd format). The red bar indicates the day of occurrence of the large EQ.

The low level preseimic signal that may exist in the recordings of figure **(7)** implies that different more sensitive methods, like the "strange attractor like" seismic precursor, must be used in order to make use of it for earthquake prediction purposes.

### 2.3. T=14days band-pass filtered data.

Now we consider a different tidal component the one of **T=14 days (M1).** The generated oscillating electric field due to the corresponding lithospheric oscillation as far as it concerns the **PYR** monitoring site is shown in figure **(8).**

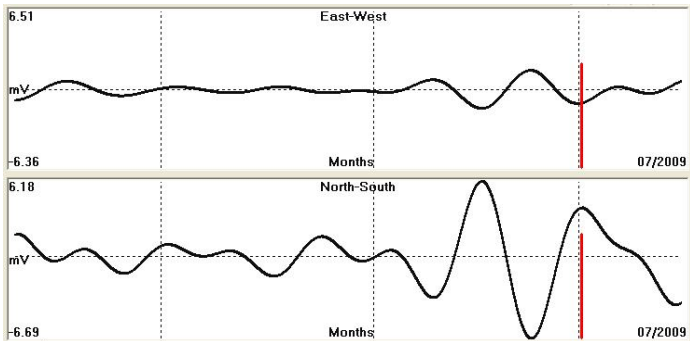

Fig.8. Band-pass filtered **(T=14days)** normalized data from **PYR** monitoring site corresponding to the period 090410 - 090715 (yymmdd format). The red bar indicates the day of occurrence of the large EQ.

The picture now is clearer. A distinct oscillation of **T=14days** started about **20** days before the seismic event. No other significant same period oscillation was observed for the entire recording period of time. Therefore we reach the conclusion that this specific oscillation was initiated some days before the EQ occurrence by the earthquake triggering mechanism.

The corresponding case for the **ATH** monitoring site is presented in figure **(9).**

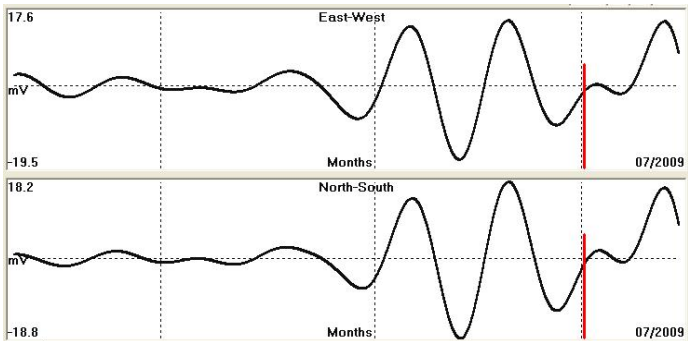

Fig.9. Band-pass filtered **(T=14days)** normalized data from **ATH** monitoring site corresponding to the period 090410 - 090715 (yymmdd format). The red bar indicates the day of occurrence of the large EQ.

The same period **(T=14days)** oscillation was observed in this case too. The difference between **ATH** and **PYR** cases is that the **ATH** anomalous signal started almost **30 days** before the seismic event and its amplitude is **3** times larger than the one at **PYR** monitoring site.

The following figure **(10)** shows the case for the **HIO** monitoring site for **T=14days**.

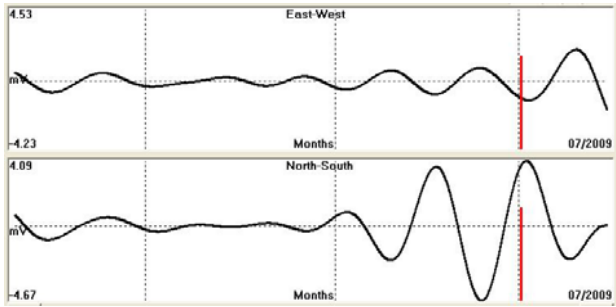

Fig.10. Band-pass filtered **(T=14days)** normalized data from **HIO** monitoring site corresponding to the period 090410 - 090715 (yymmdd format). The red bar indicates the day of occurrence of the large EQ.

Although the **HIO** monitoring site did not show any significant **T=24h** anomalous electric field, for the case of **T=14 days** behaved the same way as **PYR** and **ATH** monitoring sites. For the specific electric component of **T=14 days** all three monitoring sites were affected by the corresponding EQ triggering mechanism.

Therefore, if we compare the anomalous electric signal duration and its initiation time before the seismic event from all the previous cases that is: normalized raw data, band-pass filtered data for **T=24h** and for **T=14 days** then it is concluded that the predictive time window can be estimated at maximum as of one month.

### 2.4. Recovered true form of the generated potential.

The same raw data may be processed in a different way. Since the original recordings are made by inserting two electrodes in the ground at some distance apart, it is evident that what is originally registered is the gradient in time and distance from the epicentral area of the electric field which is generated by some physical mechanism at the focal area (Thanassoulas 2007, Thanassoulas 2008).

Consequently, if we integrate along time the original gradient data, keeping the distance from the epicetral area (and the length of the dipole) constant then the resulted potential will be a function of time of the same form as it was initially generated in the focal area but decreased in amplitude due to the existing distance between the dipole location and the focal area.

The raw data from all three monitoring sites have been processed with the earlier mentioned methodology and the results are presented in the following figures **(11, 12, 13).**

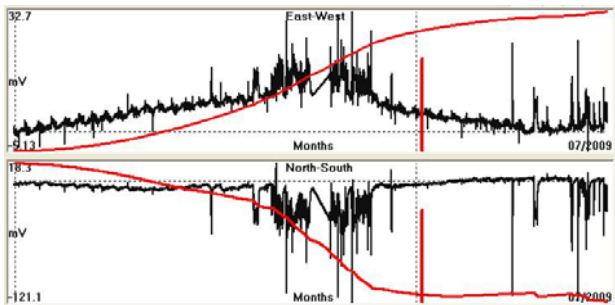

Fig.11. Normalized raw (black line) and integrated along time (red line) data for the period of time 090531-090715 (yymmdd format) from **PYR** monitoring site. The red bar indicates the day of occurrence of the large EQ.

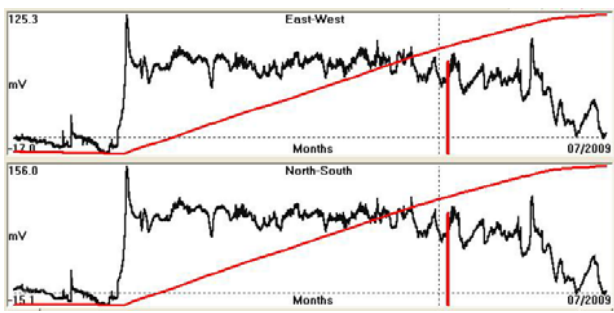

Fig.12. Normalized raw (black line) and integrated along time (red line) data for the period of time 090612 – 090708 (yymmdd format) from **ATH** monitoring site. The red bar indicates the day of occurrence of the large EQ.

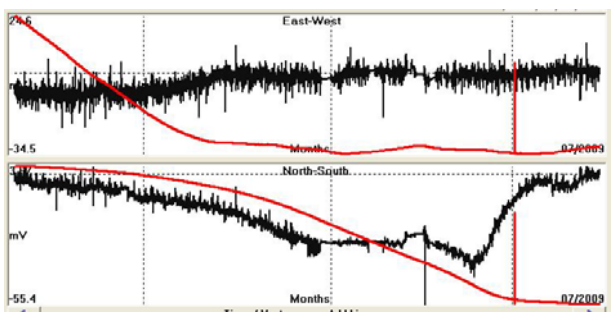

Fig.13. Normalized raw (black line) and integrated along time (red line) data for the period of time 090410-090715 (yymmdd format) from **HIO** monitoring site. The red bar indicates the day of occurrence of the large EQ.

In all three previous cases the integrated electric field exhibits the same behavior. Actually it shows the typical form of the potential generated by the deformation of a rock specimen, under increasing applied stress, provided that the rock specimen has some piezoelectric properties. The latter is the case of the lithosphere where quartzite is one of its mostly met constituents.

It must be pointed out that the large EQ occurs at the time when the rock deformation or the stress – strain relation is no more linear (Jaeger, 1974) that is the rock formation has reached the critical point of its collapse.

The case **of PYR** monitoring site **(fig. 11)** indicates that both dipoles were affected by the preseismic electric signal while at the case of **HIO** monitoring site **(fig. 13)** only the **NS** dipole was affected due to the fact that the EQ is located southern from **HIO** monitoring site. The case of **ATH** monitoring site **(fig. 12)** is not so much clear as the one in **HIO** and **PYR** monitoring sites. Nevertheless, it exhibits more or less the same form of integrated data.

This specific behavior of the integrated Earth's electric field signifies a very short time window for the time of occurrence of the pending large EQ as of a few days. Therefore, the continuous monitoring of this type of integrated data can provide valuable early warning signals before the occurrence time of the pending large EQ.

## 2.5. Determining the azimuthal direction of the Earth's electric field intensity vector.

Another method to view the registered electric data from all monitoring sites is to calculate, in each case, its azimuthal direction of the electric field intensity vector. Theoretically, and over homogeneous Earth, that vector will point towards the location of the generating mechanism. Consequently, if the very same electric field has been registered at various places, the corresponding azimuthal vectors calculated at each monitoring site will intersect at the location of the generating mechanism that is the epicenter of the large EQ (Thanassoulas 1991, 1991a). In the next figure **(14)** the azimuthal direction of the Earth's electric field intensity vector has been calculated at a time two days before the large EQ occurrence using the normalized raw data registered at **HIO** monitoring site.

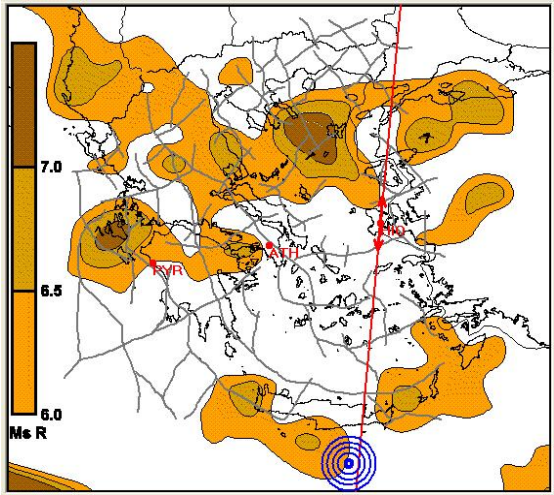

Fig. 14 Azimuthal direction of the Earth's electric field intensity vector (red line) calculated two days before the large EQ occurrence from normalized raw data registered at **HIO** monitoring site. The blue concentric circles indicate the location of the large EQ. The background map reflects the seismic potential distribution all over Greece in terms of maximum expected EQ magnitude (see scale in map). The gray thick lines represent the deep lithospheric fracture zones – faults determined by the inversion of the corresponding gravity field (Thanassoulas 1998, 2007).

The coincidence of the azimuthal direction of the intensity vector of the anomalous (preseismic) electric field calculated at **HIO** monitoring site to the location of the large EQ is remarkable. The latter indirectly suggests a rather homogeneous large scale Earth when considering such long wavelengths of the used electric field.

A generalization of this calculation is to apply it (at minute's intervals as the sampling rate is) all along the duration of the anomalous electric field and to average the obtained discrete azimuth values (Thanassoulas 2007). This operation is presented in the following figure **(15).**

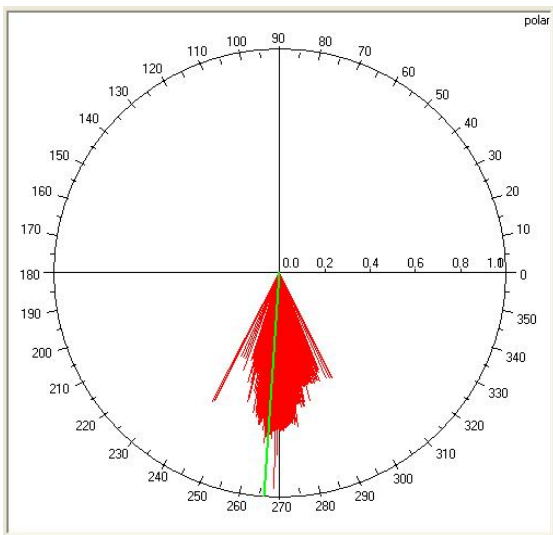

Fig.15. Azimuthal direction (red lines) of the Earth's electric field intensity vector calculated from the normalized raw data at minute's intervals all along the anomalous electric field registered at **HIO** monitoring site. The green line indicates the average azimuthal value.

It must be pointed out that when using this presentation method calculations are made for one trigonometric quadrant and the results are plotted simultaneously to the **azimuth** and **azimuth+π** directions. This is due to the fact that the true polarity of the generated field is unknown.

In the following figure **(16)** the results presented in figure **(15)** are compared to the location of the large EQ.

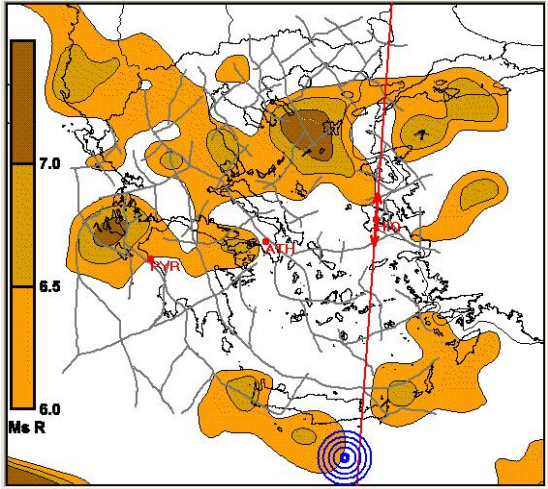

Fig. 16 Average azimuthal direction of the Earth's electric field intensity vector (red line) calculated all along the anomalous electric field before the large EQ occurrence from normalized raw data registered at **HIO** monitoring site. The blue concentric circles indicate the location of the large EQ. The background map reflects the seismic potential distribution all over Greece in terms of maximum expected EQ magnitude (see scale in map). The gray thick lines represent the deep lithospheric fracture zones – faults determined by the inversion of the corresponding gravity field (Thanassoulas 1998, 2007).

The small discrepancy observed between figures **(14)** and **(16)** is due to the fact that figure **(16)** shows the result of a large number of azimuth determinations while some part of them is of low signal to noise ratio. The latter is caused by the existing noise in the signal.

In cases when the signal of interest is heavily affected by a broad-band noise, then the data takes different filtering treatment so that the noise which is present in the band of interest can be eliminated. Such a method was introduced by Thanassoulas (2007) and Thanassoulas et al. (2008a). In this method, instead of filtering out the noise, by classical methods, from the band of interest which is rather impossible without eliminating the signal of interest, the existing noise is attacked by injecting appropriate external noise in the data aiming into canceling out only the existing noise in the band of interest and not the wanted signal. This method was applied on the normalized data recorded by **PYR** monitoring site. The best noise injection parameter value for this case is **P = 1** and the obtained results are presented in the following figure **(17).**

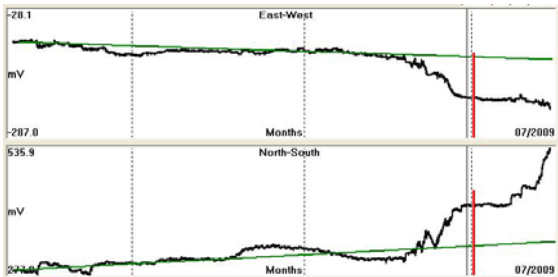

Fig.17. Noise injected **(P = 1)** normalized raw data registered from **PYR** monitoring site for the period of time 090410-090715 (yymmdd format). The red bar indicates the day of occurrence of the large EQ. The green line indicates the adopted and extracted from the data linear trend.

The azimuthal direction of the Earth's electric field registered at **PYR** monitoring site was calculated, as in the case for **HIO** monitoring site, at a time of two days before the large EQ occurrence. The determined azimuthal direction is shown in the following figure **(18).**

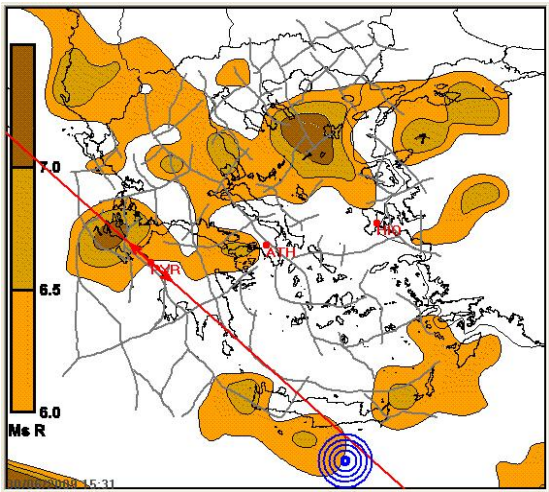

Fig.18. Azimuthal direction of the Earth's electric field intensity vector (red line) calculated two days before the large EQ occurrence from noise injected **(P=1)** normalized raw data registered at **PYR** monitoring site. The blue concentric circles indicate the location of the large EQ. The background map reflects the seismic potential distribution all over Greece in terms of maximum expected EQ magnitude (see scale in map). The gray thick lines represent the deep lithospheric fracture zones – faults determined by the inversion of the corresponding gravity field (Thanassoulas 1998, 2007).

Further more, for the case of **PYR** monitoring site, the azimuthal direction of the Earth's electric field (after noise injection) has been calculated all along the anomalous electric field **(see fig. 17)**. The obtained results are presented in the following figure **(19).**

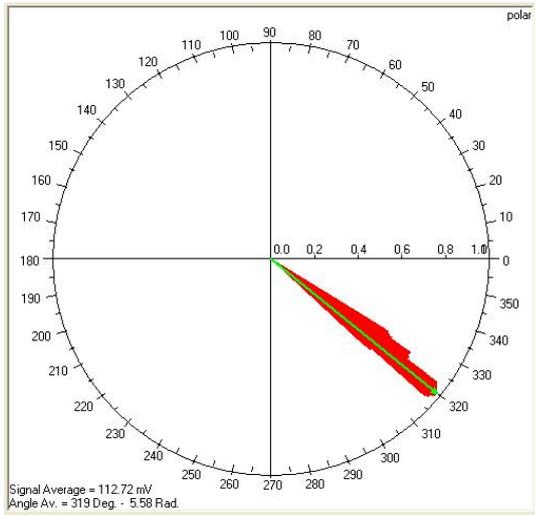

Fig.19. Azimuthal direction (red lines) of the Earth's electric field intensity vector calculated from the noise injected **(P=1)** normalized raw data at minute's intervals all along the anomalous electric field registered at **PYR** monitoring site. The green line indicates the average azimuthal value.

The very same conditions, as far as it concerns the polarity of the used electric field, hold in this case too as in figure **(15).**

At the next figure **(20)** a comparison is made between the calculated average values of the azimuthal directions to the location of the large EQ.

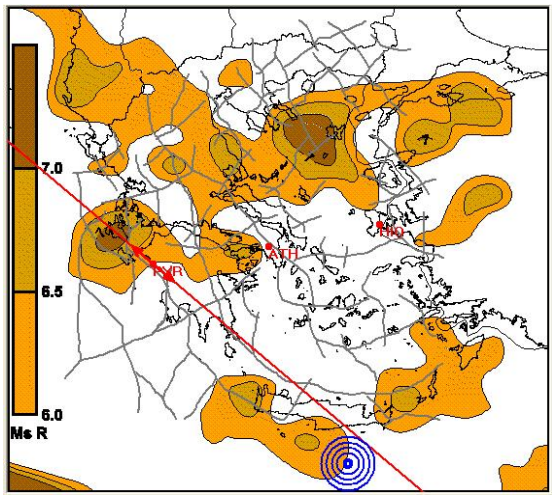

Fig.20. Average azimuthal direction of the Earth's electric field intensity vector (red line) calculated all along the anomalous electric field before the large EQ occurrence from noise injected **(P=1)** normalized raw data registered at **PYR** monitoring site. The blue concentric circles indicate the location of the large EQ. The background map reflects the seismic potential distribution all over Greece in terms of maximum expected EQ magnitude (see scale in map). The gray thick lines represent the deep lithospheric fracture zones – faults determined by the inversion of the corresponding gravity field (Thanassoulas 1998, 2007).

An immediate result of the previous calculations is the possibility to determine, with a quite acceptable accuracy, the epicenter location of the imminent large EQ. At first, an attempt is made by using the calculated azimuthal directions from figures **(14, 18).**

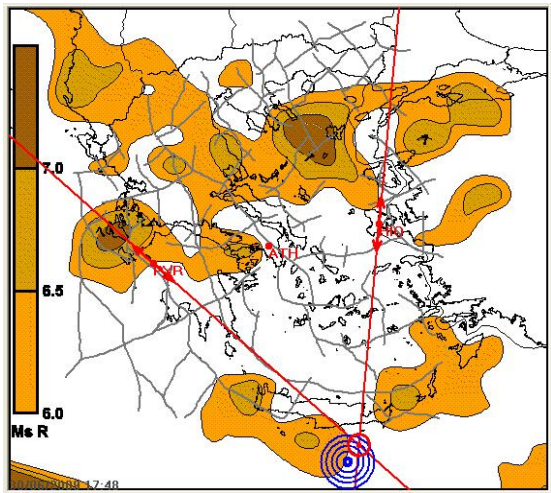

Fig. 21. Epicentral area determination, two days before its occurrence, of the large EQ, from **PYR – HIO** monitoring sites. The background map reflects the seismic potential distribution all over Greece in terms of maximum expected EQ magnitude (see scale in map). The gray thick lines represent the deep lithospheric fracture zones – faults determined by the inversion of the corresponding gravity field (Thanassoulas 1998, 2007).

The calculated (red circle) from the electric signals epicentral area differs from the seismological (concentric blue circles) one for only **37 Km** just a couple of days before the occurrence of the large EQ.

Furthermore, it is interesting to test this epicenter solution in terms of considering the calculated azimuthal direction as a time function. For this purpose instead of intersecting two distinct azimuthal directions, two angles are intersected (Thanassoulas, 2007). The first one is defined by the extreme edge values of the azimuthal directions **(fig. 19)** calculated from **PYR** monitoring site, while the second is the corresponding one from **HIO (fig. 15)** monitoring site. By using this methodology it is possible to investigate the spatial extent of possible epicenter solutions which rather suggest a regional seismogenic area within which the large EQ could occur. The latter is shown for the present case in figure **(22).**

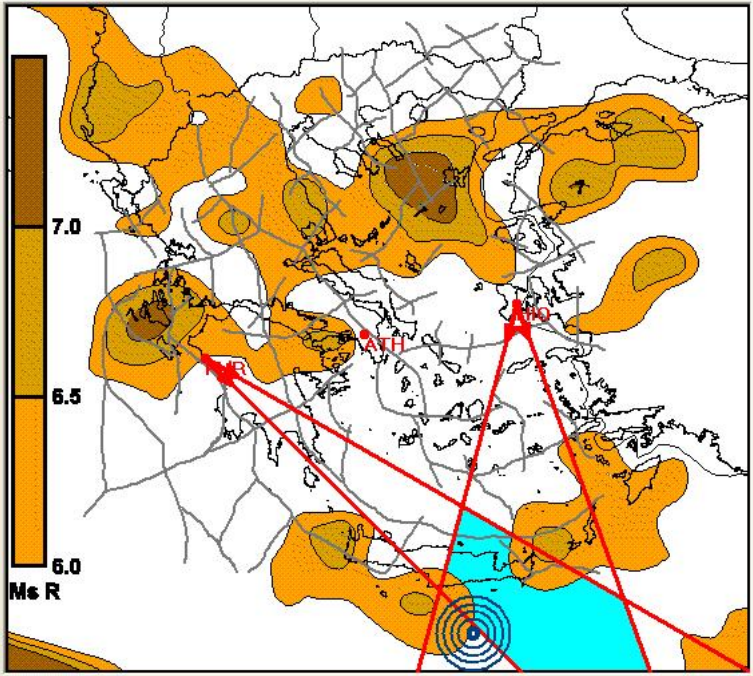

Fig. 22. Regional seismogenic area (shaded in light blue) determined from **PYR – HIO** monitoring sites. The blue concentric circles indicate the location of the large EQ. The background map reflects the seismic potential distribution all over Greece in terms of maximum expected EQ magnitude (see scale in map). The gray thick lines represent the deep lithospheric fracture zones – faults determined by the inversion of the corresponding gravity field (Thanassoulas 1998, 2007).

The defined with this methodology area (shaded light blue) prescribes the spatial distribution of possible epicentral areas which satisfy both angle ranges calculated for **PYR** and **HIO** monitoring sites. From figure **(22)** it is shown that the studied EQ occurred at the margins of this area.

## 2.6. Application of the "strange attractor like" seismic electric precursor.

In this part of the paper the Earth's electric field is simultaneously investigated at pairs of related electric signals formed from any possible combination of two monitoring sites out of three **(PYR, ATH and HIO)**. For this purpose the method of the "strange attractor like" has been applied on the monochromatic and of **T=24h** electric signals obtained from each monitoring site (Thanassoulas et al. 2008b, 2008c, 2009, 2009a).

The method has been applied for a period of time that spans from June 21$^{st}$, 2009 to July 2$^{nd}$, 2009. Actually, the "strange attractor like" electric precursor behavior is investigated for a period of ten **(10)** days before, the day during the large EQ occurrence and one immediately after it. The calculated "strange attractor like" precursor is presented in groups of **(12)** daily maps each corresponding to a pair of monitoring sites as follows in figures **(23, 24, 25).** The corresponding date of each map is marked at its lower left corner.

**"Strange attractor like" seismic electric precursor: ATH – PYR from 090621 to 090702 (yymmdd format) from left to right.**

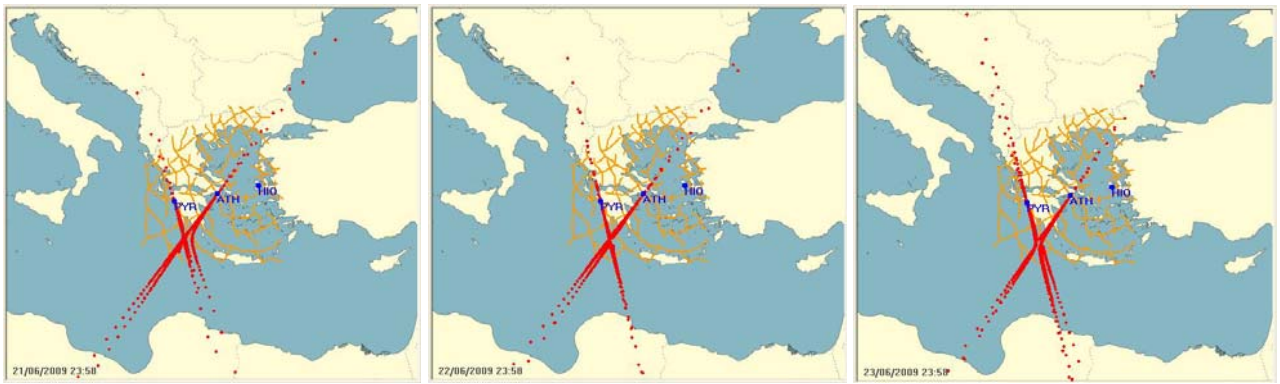

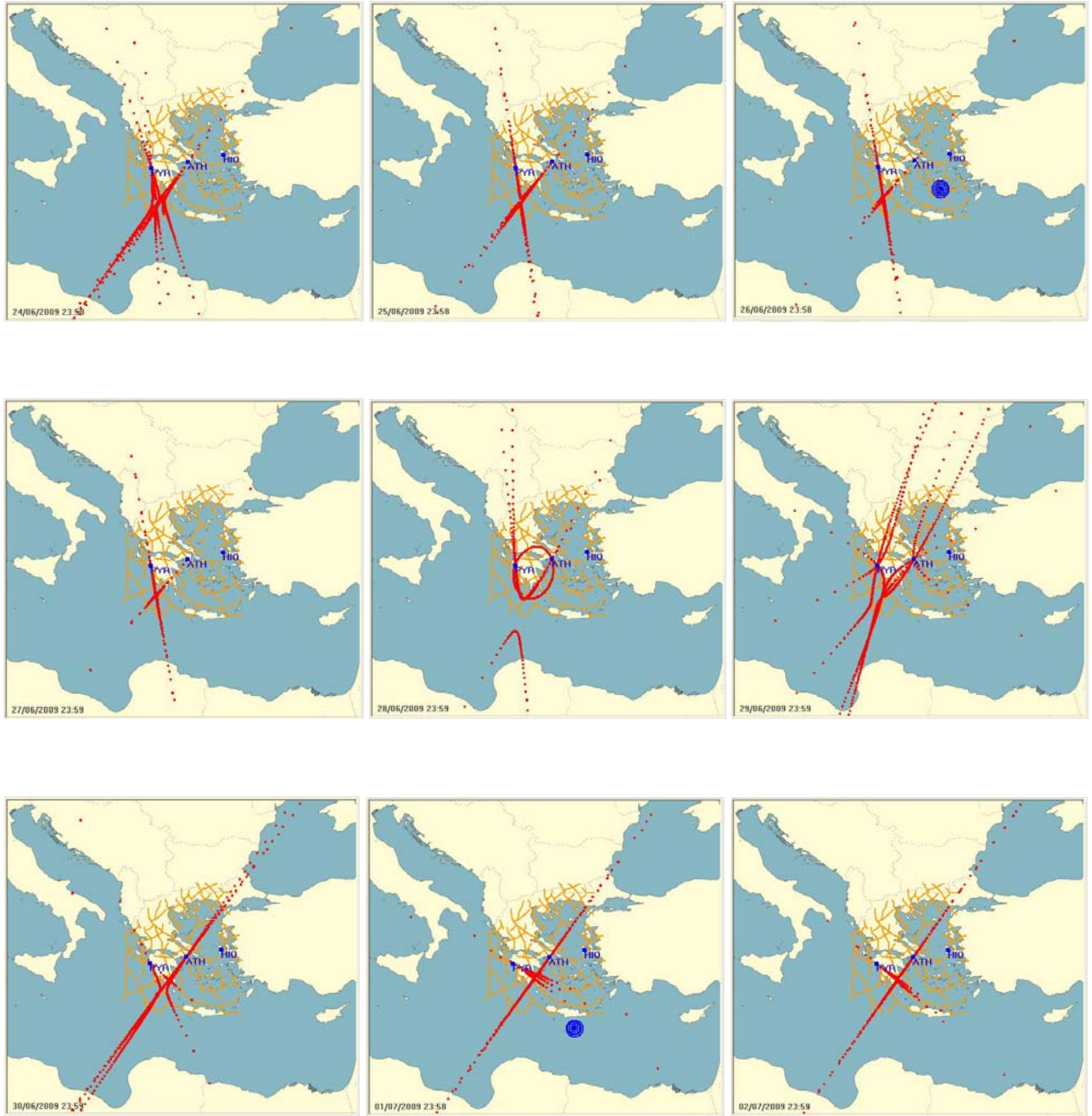

Fig.23. "Strange attractor like" seismic precursor (red dotted lines) calculated for the period of time from June 21st, 2009 to July 2nd, 2009 from the oscillating component **(T=24h)** of the Earth's electric field obtained from **PYR** and **ATH** monitoring sites. The blue concentric circles correspond to: June 26th, 2009 EQ magnitude **Ms = 5.5R**, July 1st, 2009 EQ magnitude **Ms = 6.3R**.

From the inspection of the presented maps of figure **(23)** it is shown that the "strange attractor like" seismic precursor developed on the 28th of June 2009 that is three **(3)** days before the large EQ occurrence and lasted for one **(1)** day.

**"Strange attractor like" seismic electric precursor: ATH – HIO from 090621 to 090702 (yymmdd format) from left to right.**

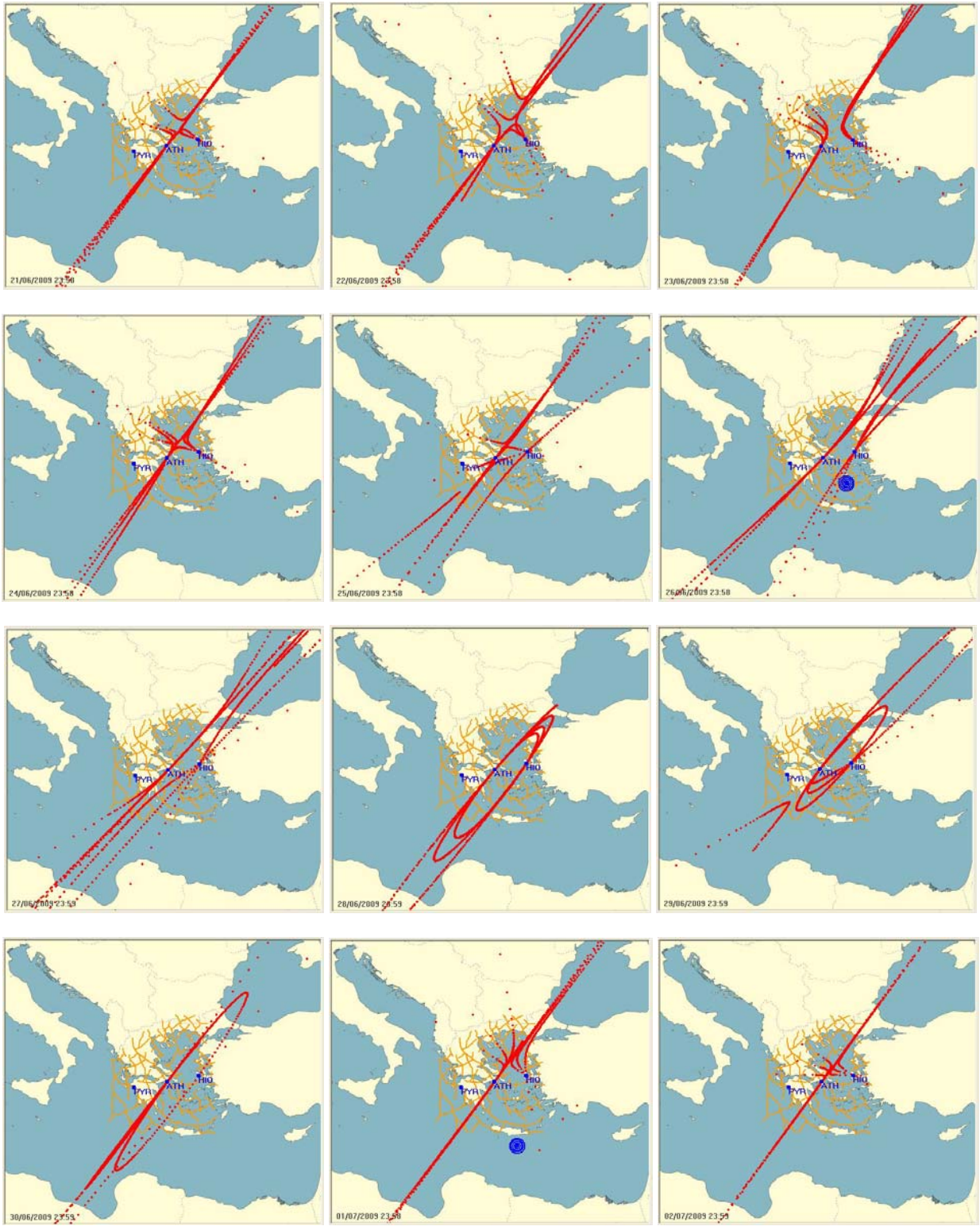

Fig. 24. . "Strange attractor like" seismic precursor calculated for the period of time from June 21st, 2009 to July 2nd, 2009 from the oscillating component **(T=24h)** of the Earth's electric field obtained from **ATH** and **HIO** monitoring sites. The blue concentric circles correspond to: June 26th, 2009 EQ magnitude **Ms = 5.5R**, July 1st, 2009 EQ magnitude **Ms = 6.3R**.

From the inspection of the presented maps of figure **(24)** it is shown that the "strange attractor like" seismic precursor developed on the 28[th] of June 2009 that is three **(3)** days before the large EQ occurrence and lasted for three **(3)** days.

**"Strange attractor like" seismic electric precursor: PYR – HIO from 090621 to 090702 (yymmdd format) from left to right.**

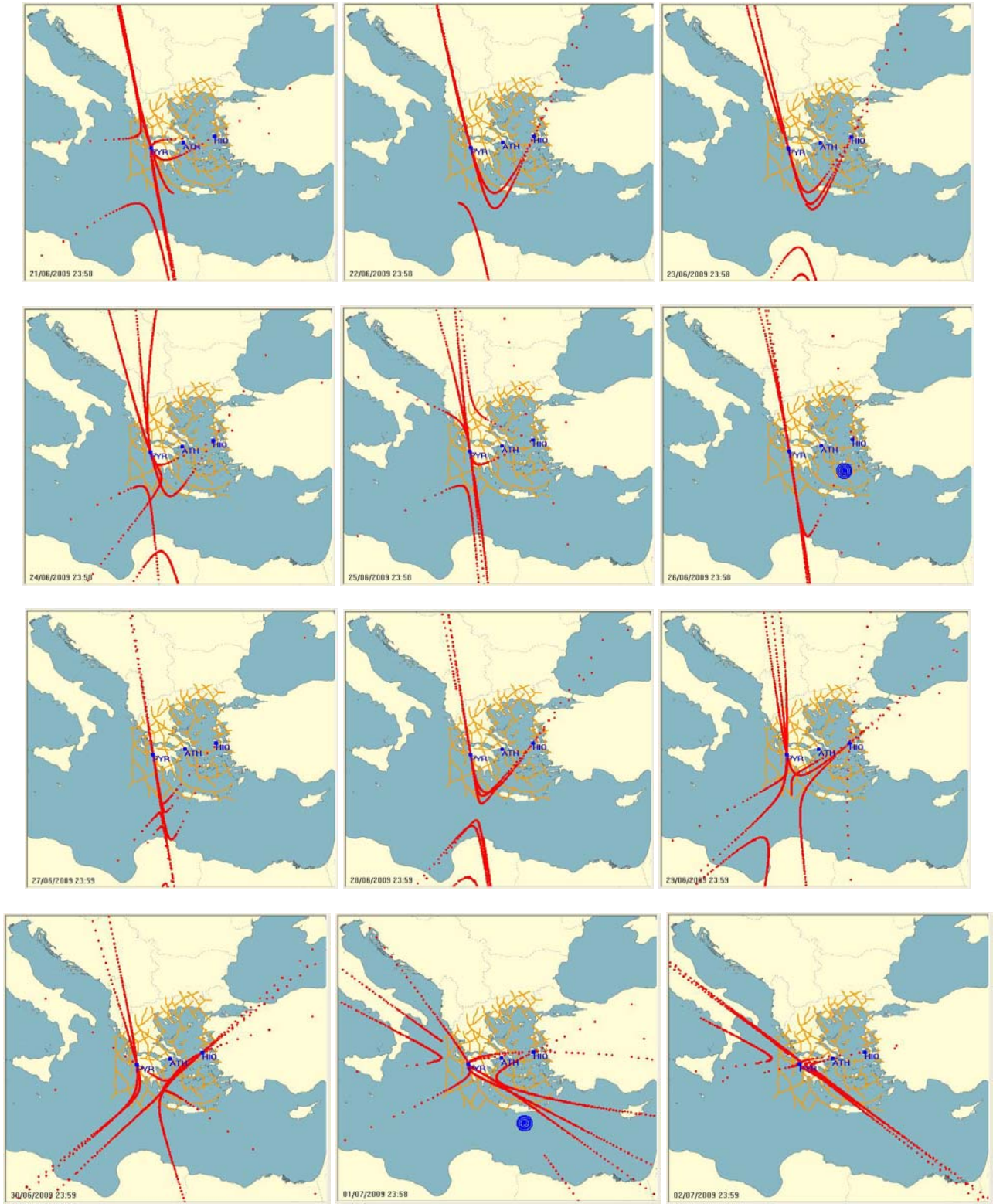

Fig. 25. "Strange attractor like" seismic precursor calculated for the period of time from June 21[st], 2009 to July 2[nd], 2009 from the oscillating component **(T=24h)** of the Earth's electric field obtained from **PYR** and **HIO** monitoring sites. The blue concentric circles correspond to: June 26[th], 2009 EQ magnitude **Ms = 5.5R**, July 1[st], 2009 EQ magnitude **Ms = 6.3R**.

From the inspection of the presented maps of figure **(25)** it is shown that the "strange attractor like" seismic precursor developed concurrently with the large EQ occurrence day and lasted for one **(1)** day.

An interesting observation that can be made from figures **(23, 24)** is the fact that the "strange attractor like" electric precursor was initiated at the same day. Moreover, that precursor for the specific pairs of monitoring sites diminished before the large EQ occurrence. This behavior of it complies with the one reported by Thanassoulas et al. (2009). The absence of the "strange attractor like" precursor between the **PYR** and **HIO** monitoring sites before the large EQ occurrence could be explained by the "loose" electric preseimic signal coupling that is probably due to the large distance observed between them. The concurrent presence of the "strange attractor like" precursor, for the same monitoring sites, with the large EQ occurrence day could be explained by the fact that a day before the **M1** tidal wave reaches its largest peak value, probably, the lithosphere just reached the critical conditions for generating an electric signal of such amplitude capable to trigger both **PYR** and **HIO** monitoring sites into generating a "strange attractor like" precursor. It must be pointed out that the raw "strange attractor like" seismic precursor can be decomposed into simpler components (Thanassoulas et al. 2009b) and therefore, the way this precursor is generated and related to the activated regional seismogenic area is not yet very clear. The postulated model by Thanassoulas et al. (2009b) assumes the simultaneous presence of two distinct activated current sources (seismogenic areas) which arises the question whether one of them could be the pending large EQ focal area. The latter needs definitely more detailed analysis.

### 3. Conclusions

The analysis of the registration of the Earth's electric field by three monitoring sites **(PYR, ATH** and **HIO)** prior to the occurrence of the large EQ of July $1^{st}$, 2009 revealed the existence of various seismic electric precursors. These electric precursors can be grouped as follows:

   a. **Electric precursors observed at the normalized raw dada** such as: **group of sharp electric pulses** that follow the tertiary creep model reported by Morgounov (2001), **plateau-like anomalous electric field** complying to the piezoelectric model reported by Thanassoulas (1991, 2007) and **Very Long Period (VLP) anomalies** as observed in the past by Sobolev (1975). All these types of anomalous electric field suggest a time window for the occurrence of the pending large EQ of the order of some days. Actually, the electric pulses are generated at the last phase of the rock deformation just before fracturing takes place. The plateau-type anomalous field indicates that the last phase, before the rock fracturing, has been reached and it is indicated by the generation of that characteristic form of the anomalous field due to the stress – strain relation that holds for the focal area rock formation. The **VLP** anomalous field steep rising brunch of it, at its end, is the time window, in the majority of the cases, where the large EQ takes place. Consequently, these three different anomalous electric preseismic signals can be combined into a time window that will satisfy all of them.

   b. **Electric precursors observed after band-pass filtering of the normalized raw data** such as: **increased amplitude monochromatic (T=24h) oscillating electric field** (Thanassoulas et al. 1993) obtained after band-pass filtering of the normalized raw data. This electric precursor is generated by the lithosphere tidally triggered oscillation due to **K1** and **P1** components of the tidal wave. A **similar oscillating (T=14 days) electric field** is generated when the lithosphere is triggered by the **M1** tidal wave component. In this case it seems that the **M1** component of the tidal wave was the primary trigger of the lithosphere oscillation. These oscillating components are mostly generated when the focal area has reached critical conditions before failure. In such a case the tidal wave works as the final needed trigger to initiate the EQ. A time window of a few days can be estimated from both these precursors.

   c. **Electric precursors observed in the integrated along time normalized raw data.** In this case the original form of the generated in the focal area electric potential is obtained. That potential represents the typical stress – potential form function of the piezoelectric mechanism or any other mechanism that behaves as a piezoelectric one, thus being capable to identify the time window within a few days in which the rock formation of the focal area will fracture.

   d. **The "strange attractor like" precursor.** In this case the interrelationship of signals recorded by two different monitoring sites due to "strong stress coupling?, not yet clear" is revealed by the generation of the "strange attractor like" electric precursor, thus suggesting the last phase of the triggering of a large EQ has been reached. The determined time window for the occurrence of a large EQ is of the order of a few days.

Moreover, it is possible to further process the normalized raw data so that the epicenter area of the pending large EQ can be estimated quite accurately or in a regional form by calculating the azimuth direction of the Earth's electric field intensity vector at the different used monitoring sites and finally using simple triangulating techniques.

In conclusion, the registered Earth's electric field data incorporate valuable information about the time window of the occurrence and the location of a pending large EQ. What is really needed to extract this information is some simple physics methodologies applied on a well justified physical model. `